\documentclass[amsmath,amssymb,preprintnumbers,prc,showpacs,showkeys,byrevtex]{revtex4}
\usepackage{graphicx}
\begin{document}
\preprint{CERN-PH-TH/2005-210}
\title{
Comparison of Chemical Freeze-Out Criteria \\
in Heavy-Ion Collisions}
\author{J. Cleymans}
\affiliation{UCT-CERN Research Centre and Department  of  Physics,
University of Cape Town, Rondebosch 7701, South Africa}
\author{H. Oeschler}\affiliation{Darmstadt University of Technology, D-64289 Darmstadt, Germany\\
UCT-CERN Research Centre and Department  of  Physics, \\
University of Cape Town, Rondebosch 7701, South Africa}
\author{K. Redlich}\affiliation{Institute of Theoretical Physics, University of Wroc{\l}aw,\\
 Pl. Maksa Borna 9, 50-204  Wroc{\l}aw, Poland\\
CERN TH, CH 1211 Geneva 23, Switzerland}
\author{S. Wheaton}\affiliation{
UCT-CERN Research Centre and Department  of  Physics, \\
University of Cape Town, Rondebosch 7701, South Africa\\
Darmstadt University of Technology, D-64289 Darmstadt, Germany
}
\date{\today}
\newcommand{\eovern}{\left<E\right>/\left<N\right>}
\begin{abstract}
One of the most remarkable results to emerge from heavy-ion
collisions over the past two decades is the  striking regularity
shown by particle yields at all energies. This has led to several
very successful proposals describing particle yields over a very
wide range of beam energies, reaching from 1 A GeV up to 200 A
GeV, using only one or two parameters. A systematic comparison of
these proposals is presented here. The conditions of fixed  energy
per particle, baryon+anti-baryon density, normalized entropy
density  as well as percolation model are investigated. The
results are compared with the most recent chemical freeze-out
parameters obtained in the thermal-statistical  analysis of
particle yields. The  sensitivity and dependence of the results on
parameters is analyzed and discussed. It is shown that in the
energy range above the top  AGS energy, within present accuracies,
all chemical freeze-out criteria give a fairly good description of
the particle yields. However, the low energy heavy-ion data favor
the constant energy per particle as a unified condition of
chemical particle freeze-out. This condition also shows the
weakest sensitivity on model assumptions and  parameters.
\end{abstract}
\pacs{24.10.Pa,25.75.Dw,12.38.Mh} \keywords{chemical, freeze-out,
heavy-ion collisions, hadron gas} \maketitle
\section{Introduction}
One of the most  remarkable results to  emerge from relativistic
heavy-ion collisions over the past years  is the  striking
regularity shown by particle yields at all beam energies. From the
lowest SIS  up to the highest RHIC energies, all results on
particle multiplicities are consistent with the assumption of
chemical equilibrium in the final-state fireball produced after
heavy-ion impact~\cite{review}. The particle yields are found to
be described, with remarkable precision, by a thermal-statistical
model that assumes approximate chemical equilibrium
\cite{review,re1,re2,re3,kaneta2,magestro,manninen,oeschler59,becattini1,averbeck,ingrid,larissa,star,broniowski,bron-1}.
For a given collision energy, the thermal-statistical model with
only two parameters,  the temperature $(T)$ and baryon chemical
potential $(\mu_B)$, provides a very systematic  description of
particle yields.

The partition function used in heavy-ion collisions has also been
shown to be consistent with results obtained from lattice gauge
theory (LGT)  in the hadronic phase \cite{lgt,lgt1}. This,
together with the phenomenological success of the
thermal-statistical model, indicates that the resonance gas
description of the hadronic phase is a remarkably accurate
approximation of QCD thermodynamics in  the hadronic confined
phase.

Using  this description as a starting point, a further interesting
systematic behavior of thermal parameters has emerged from
particle yields in  heavy-ion collisions from SIS up to RHIC. With
increasing collision energy, there is an increase of the chemical
freeze-out temperature, $T$, and a corresponding decrease of the
baryon chemical potential, $\mu_B$ . In the
{$(T,\mu_B)$}~--{~plane} the freeze-out parameters lie on a
curve connecting the lowest SIS through AGS, SPS and RHIC points, 
with a temperature at $\mu_B=0$ that corresponds to the critical
temperature expected for deconfinement in LGT
\cite{review,stachel,krtc}. Such a regular behavior of the
freeze-out parameters has called for an interpretation 
based on the interaction dynamics. There are properties of the thermal fireball at
chemical freeze-out that are common to all collision energies.
Such common properties provide  unified chemical freeze-out
conditions in heavy-ion collisions at all energies.

The knowledge of such   unified freeze-out properties is
useful since:
\begin{itemize}
\item they make it possible to determine particle excitation
functions and to study their systematic properties \cite{pbmco}.
Consequently,  general conclusions can be drawn about  the energy
dependence of particle production in heavy-ion collisions; \item
the freeze-out conditions are relevant for a successful
description of  particle spectra;
\item on the theoretical level the unified freeze-out conditions
are relevant for the  understanding of  particle production
dynamics. The chemical freeze-out line  has been  interpreted as
the inelasticity line in heavy-ion collisions \cite{uh}, as it
separates the stage where the evolution dynamics of the collision
fireball is dominated by inelastic instead of  elastic processes.
\end{itemize}

The first unified chemical freeze-out  criterion  was recognized
\cite{redlich1,redlich2} when comparing the thermal parameters at
SIS energy with those obtained at SPS.  It was shown that the
average energy per particle  at SIS energy reaches  approximately
the same value of 1 GeV as calculated at the critical temperature
expected for deconfinement at $\mu_B=0$. In addition, known
results for chemical freeze-out parameters at the AGS also
reproduced the same value of energy per particle.  Thus, it was
suggested that the condition of a fixed energy per hadron is the
chemical freeze-out criterion in heavy-ion collisions. More
precise determinations prefer a value $\eovern\approx$ 1.08 GeV.

%In the last two years  a whole set of new experimental data at SPS
%from the  NA49 and CERES collaborations  as well as at RHIC were obtained.
%These data were  analyzed within
%the thermal-statistical model to determine the chemical freeze-out parameters.
%All these new results provide further support for the freeze-out
%condition proposed in
%\cite{redlich1,redlich2}.

In addition to the fixed  $\eovern$ criterion,  alternative
proposals have been made  to describe chemical freeze-out in
heavy-ion collisions at all
energies~\cite{pbm,satz,stankiewicz,horn,steinberg,tawfik1,tawfik2}:
\begin{itemize}
\item a fixed value for the sum of
baryon and anti-baryon densities, $n_B+n_{\bar{B}}$, of
approximately 0.12/fm$^3$~\cite{pbm};
\item a self-consistent
equation for the densities based on geometric estimates using
percolation theory~\cite{satz};
\item a fixed value of the entropy
density, $s/T^3$, of approximately
7~\cite{stankiewicz,steinberg,horn,tawfik1,tawfik2}.
\end{itemize}
All of these proposals  have been used with considerable success, and
it is the purpose of the present article to
study and compare these different criteria for particle freeze-out to
constrain their validity on the basis of  available
experimental data and corresponding  statistical model analyses.
It
is not always straightforward to judge their quality, since
small variations in parameters
can sometimes lead to substantial shifts in results.
In our
analysis, we quantify the  explicit dependence of different freeze-out
criteria on excluded volume corrections, the resonance composition
of the fireball, strangeness content   and the corrections
related to the canonical implementation  of strangeness
conservation.
In particular, excluded volume corrections~\cite{hardcore,gorenstein}
have a very strong impact on the validity of most of the criteria.

We  show that in the energy range above the top  AGS beam energy
and within present accuracies, it is rather difficult to
distinguish between different freeze-out conditions. However, the
lower energy heavy-ion data favor the constant energy per particle
as the unified condition of chemical particle freeze-out. It also
exhibits  the least sensitivity to model assumptions and the most
stability
 to small changes in the values of the  parameters.
\section{Global Description}
\begin{figure*}[t]
\begin{center}
\includegraphics[width=12cm]{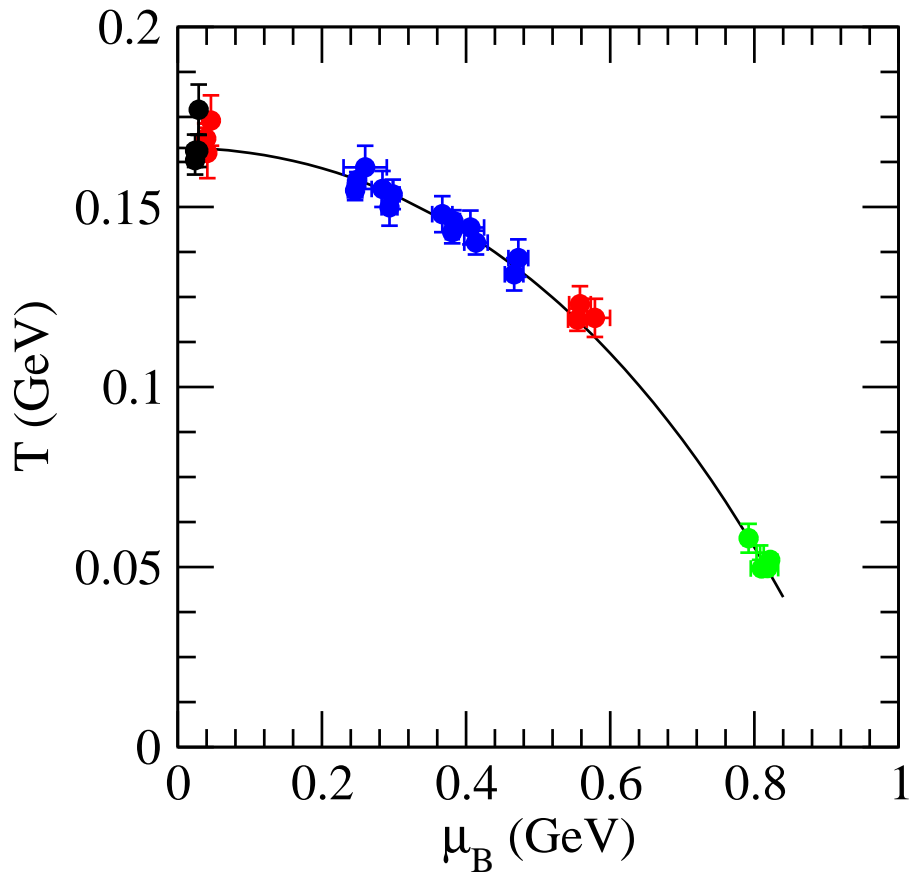}
\end{center}
\caption{ Values of $\mu_B$ and $T$ for different energies. The solid line is a parameterization corresponding to
$T(\mu_B) \approx 0.17 - 0.13\mu_B^2 -0.06 \mu_B^4 $ (see text).  } \label{pdg}
\end{figure*}

The results from particle yields obtained  by different groups for
the thermal parameters are
not always identical. The differences are small and
never exceed  the level of a few percent. Nevertheless, such
differences  play a role when one tries to determine unified
descriptions over a wide range of beam energies. The origin of these
differences has not been fully  studied until now. However, to a large
extent these differences could be attributed to different
assumptions on the number of  resonances, their widths,
the treatment of weak decays,
the description of strangeness saturation
and  how, or if at all,  repulsive interactions have been
included in the calculations. To pin down the differences
 would involve  a very detailed knowledge of
programming codes. This is not yet possible, as codes are not
always made available. In order to have a consistent presentation,
we have made some choices which are biased by our own
experience: the code used in this paper is
available for inspection~\cite{thermus}. Our analysis relies on
the most recent results obtained in statistical-thermal model fits
to Au+Au and Pb+Pb systems, performed by numerous groups over a
wide range of
energies~\cite{becattini1,magestro,kaneta2,manninen,oeschler59,averbeck,ingrid,larissa,star,broniowski,bron-1}.
These results are summarized in Table~\ref{tab:ModelFits} and are also included in Figure~\ref{pdg}.\\

\begin{table}
\begin{tabular}{|c|c|r@{$\pm$}l|r@{$\pm$}l|c|}\hline
Collision System and Energy & Ref. & \multicolumn{2}{|c|}{$T$}   & \multicolumn{2}{|c|}{$\mu_B$} & Include in       \\
       &      & \multicolumn{2}{|c|}{(MeV)} & \multicolumn{2}{|c|}{(MeV)}   & Fits \\\hline\hline
\multicolumn{7}{|c|}{RHIC}\\\hline\hline Au+Au $\sqrt{s}=200$ AGeV
& \cite{magestro} & 177 & 7 & 29 & 6 &  \\
& \cite{star} & 163 & 4 & 24 & 4 &
\\
& \cite{bron-1} & 165.6 & 4.5 & 28.5 & 3.7 & \\\hline\hline
Au+Au $\sqrt{s}=130$ AGeV & \cite{kaneta2} & 169 & 6 & 38.1 & 4.2 & \checkmark \\
 & \cite{magestro} & 174 & 7 & 46 & 5 & \checkmark\\
& \cite{broniowski} & 165 & 7 & 41 & 5 & \checkmark \\\hline\hline
\multicolumn{7}{|c|}{SPS}\\\hline\hline
Pb+Pb 158AGeV $\sqrt{s}=17.3$ AGeV & \cite{manninen} & 157.5 & 2.2 & 248.9 & 8.2 & \checkmark \\
 & \cite{manninen} & 154.6 & 2.7 & 245.9 & 10.0 & \checkmark \\
 & \cite{ingrid} & 161.0 & 6.0 & 260.0 & 30 & \checkmark \\\hline
Pb+Pb 80AGeV $\sqrt{s}=12.3$ AGeV & \cite{manninen} & 153.5 & 4.1 & 298.2 & 9.6&\checkmark  \\
 & \cite{manninen} & 149.9 & 5.1 & 293.8 & 11.0 & \checkmark \\
 & \cite{larissa} & 155.0 & 5.0 & 284.0 & 15.0 & \checkmark \\\hline
Pb+Pb 40AGeV $\sqrt{s}=8.77$ AGeV & \cite{manninen} & 146.1 & 3.0 & 382.4 & 9.1 & \checkmark \\
 & \cite{manninen} & 143.0 & 3.1 & 380.8 & 8.9 & \checkmark \\
 & \cite{larissa} & 148.0 & 5.0 & 367.0 & 14.0 & \checkmark \\\hline
Pb+Pb 30AGeV $\sqrt{s}=7.62$ AGeV & \cite{manninen} & 140.1 & 3.3&413.7 &16.3& \checkmark\\
 & \cite{manninen} & 144.3 & 4.7&406.0 &19.1&\checkmark \\
\hline\hline
Pb+Pb 20AGeV $\sqrt{s}=6.27$ AGeV & \cite{manninen} & 131.3 & 4.5&466.7 &12.9&\checkmark \\
 & \cite{manninen} & 135.8 & 5.2&472.5 &13.7&\checkmark \\
\hline\hline
\multicolumn{7}{|c|}{AGS}\\\hline\hline
Au+Au 11.6AGeV $\sqrt{s}=4.86$ AGeV & \cite{manninen} & 118.7 & 3.1 & 554.4& 13.0 &\checkmark \\
 & \cite{manninen} & 119.2 & 5.3 & 578.8 & 15.4 & \checkmark \\
 & \cite{larissa} & 123.0 & 5.0 & 558.0 & 15.0 & \checkmark \\\hline\hline

\multicolumn{7}{|c|}{SIS}\\\hline\hline
Au+Au 1.0AGeV $\sqrt{s}=2.32$ AGeV & \cite{oeschler59} & \multicolumn{2}{|c|}{52$\pm 1.5$} & \multicolumn{2}{|c|}{822} &  \checkmark \\
 & \cite{becattini1} & 49.6 & 1 & 810 & 15 & \checkmark \\
 & \cite{becattini1} & 49.7 & 1.1 & 818 & 15 & \checkmark \\
 & \cite{averbeck} & 58 & 4 & 792 & 7 & \checkmark \\\hline
Au+Au 0.8AGeV $\sqrt{s}=2.24$ AGeV & \cite{averbeck} & 54 & 2 & 808 & 5 & \checkmark \\\hline
\end{tabular}
\caption{\label{tab:ModelFits}Results obtained in statistical-thermal model fits to Au+Au and Pb+Pb collision systems by
numerous groups over a wide range of energies. The checked entries have been included in
the fits to determine the $T(\mu_B)$ and $\mu_B(\sqrt{s})$ parameterizations
(Equations~(\ref{Eqn:T(muB)}) and (\ref{Eqn:MuB(s)}), respectively).}
\end{table}

\begin{figure*}[b]
\begin{center}
\includegraphics[width=12cm]{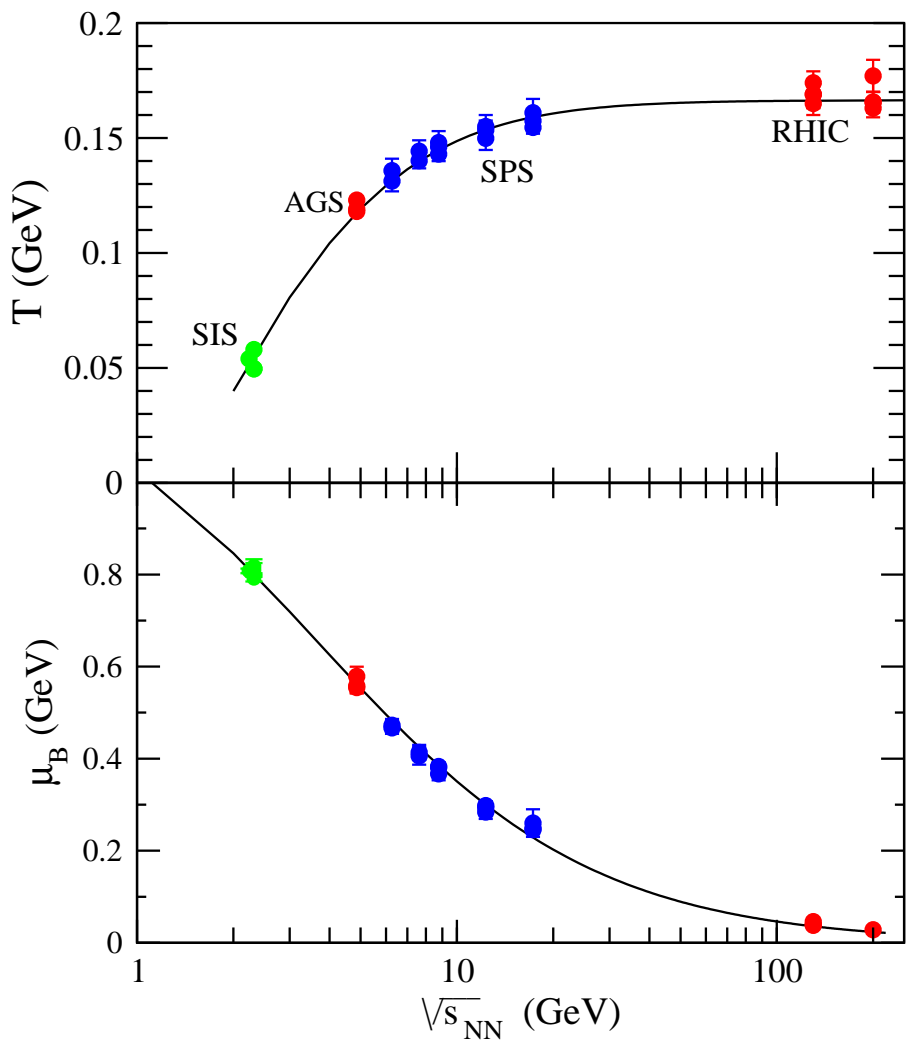}
\end{center}
\caption{
Energy dependence of the chemical freeze-out parameters $T$ and $\mu_B$. The curves
have been obtained using a parameterization discussed in the text.
}
\label{mub}
\end{figure*}

Before discussing  regularities in Table~\ref{tab:ModelFits}, we
present  first a  polynomial fit using the checked entries in
Table~\ref{tab:ModelFits} as input. This gives,
\begin{equation}
T(\mu_B) = a - b\mu_B^2 -c \mu_B^4 .
\label{Eqn:T(muB)}
\end{equation}
where $ a =  0.166 \pm 0.002$ GeV, $b = 0.139 \pm 0.016$ GeV$^{-1}$ and 
$c = 0.053 \pm 0.021$ GeV$^{-3}$.
The $\sqrt{s}=200$ AGeV points were  not included in the determination of Eq.~(\ref{Eqn:T(muB)}).
The comparison of the above equation with
the $\sqrt s$-dependent  chemical potential from
 Table~\ref{tab:ModelFits}  is shown in Figure~\ref{pdg}.
It is also
 straightforward
 to verify that at $\sqrt s=200$ GeV Eq.~(\ref{Eqn:T(muB)})
 is consistent within errors with the expected results from
 thermal model analyses.
The energy dependence of the baryon chemical potential can be
parameterized  as~\cite{wheaton}:
%\begin{equation}
%\mu_B(\sqrt{s}) = \frac{1.31~\mathrm{GeV}}{1 +
%0.276~{\mathrm{GeV}}^{-1}\sqrt{s}}, \label{Eqn:MuB(s)}
%\end{equation}
\begin{equation}
\mu_B(\sqrt{s}) = \frac{d}{1 + e\sqrt{s}}, \label{Eqn:MuB(s)}
\end{equation}
with $d = 1.308\pm 0.028$ GeV and  $e = 0.273 \pm 0.008$ GeV$^{-1}$.
The temperature $T$ and the baryon chemical potential
 $\mu_B$ are shown in Figure~\ref{mub} as a function
of the beam energy.
It provides a good quantitative
 description of the corresponding
thermal model results.
Eqs.~(\ref{Eqn:T(muB)})  and
(\ref{Eqn:MuB(s)}) lead to the  energy
dependence of
the chemical freeze-out temperature and 
baryon chemical potential shown in Figure~\ref{mub}.
In heavy ion collisions at the LHC Eq.~(\ref{Eqn:MuB(s)}) leads 
to a value $\mu_B\approx 1$ MeV.  \\
The parameterization given above should of course be
used with caution, as small changes in the parameters lead to a significant
change in the energy dependence, since particle yields depend on 
$\mu_B$ and $T$  exponentially. A particularly

%There are data obtained in heavy-ion collisions at AGS and SPS for
%2,4,6,8 and 20 AGeV which have not yet been  analyzed in terms of
%the statistical model. This is mainly because the number of
%available observables is not sufficient to constrain the thermal
%parameters. However, in view of the systematic parameterization of
%the $\sqrt s$ dependence of $T$ and $\mu_B$ in Eqs.~(\ref{Eqn:T(muB)}) 
%and (\ref{Eqn:MuB(s)}) one can calculate these
%parameters
% for the above energies and check the
%consistency of the statistical model results with the experimental
%data.
%
%
%In the following we  will use the freeze-out parameters from Table
%1 together with that from   Table 1a
% to discuss condition with the results from Table 1.}

Having established  a basic parameterization of freeze-out
conditions at different energies, we  now proceed to compare
various unified conditions  for the chemical freeze-out curve. In
the next section we will discuss how the freeze-out criteria
mentioned in the introduction live up to a comparison with the
description presented here. We will, in particular, discuss the
dependence on parameters such as the strangeness    suppression factor
$\gamma_s$, canonical corrections to strangeness conservation, the
cut-off of the hadronic  mass spectrum and, finally, excluded
volume effects.
Clearly, a superficial examination shows that many  criteria can be
remarkably effective at describing chemical freeze-out in the
restricted  energy range. The largest discrepancies, as it will be
argued, appear at low energies. However, a  closer examination
reveals that some criteria  are more robust than others when small
changes in parameters are allowed.
\section{Description of Freeze-out Criteria}

All  criteria considered here are based on the hadron resonance
gas model that was successfully used to describe particle yields
and their ratios in heavy-ion collisions. Although this model is
well known in the literature, for completeness we will summarize
the basic ideas and formulae required to  quantify each freeze-out
condition.

The model is defined by the partition function  of a resonance gas
with free particle dispersion relations for all constituents
\cite{review}. The resulting observables, however,  account for
interactions between hadrons, to the extent that the
thermodynamics of an interacting system of elementary hadrons can
effectively be approximated by a mixture of ideal gases of
stable particles and resonances. Thus, the thermodynamical
observables are  expressed as a sum over all mesonic and baryonic
degrees of freedom, as well as their known resonances.

For our further discussion, the relevant quantities are: the total
particle density $n$, total density of baryons $n_B+n_{\bar B}$, as
well as  energy $\epsilon$ and net baryon number $n_B - n_{\bar B}$ densities. All
of these observables are  functions of temperature $T$ and chemical
potentials $\mu_B,\mu_Q$ and $\mu_S$ related to conservation of
baryon number, electric charge and strangeness, respectively. The last two potentials are fixed from
the initial conditions. Thus, we are left with only two
independent variables: $T$ and $\mu_B$. The results for all relevant 
quantities obtained from
the hadron resonance gas partition function are as follows:

\noindent
 The  particle density is given by,
\begin{equation}\label{1}
n= \sum_i g_i \int\frac{d^3p}{(2\pi)^3} f_i,
\end{equation}
the entropy density by,
\begin{equation}
\label{Entropy}
s = \sum_i\frac{g_i}{2\pi^2} \int_0^\infty p^2dp \left[ - f_i\ln f_i \mp (1 \mp f_i)\ln( 1 \mp f_i)  \right],
\end{equation}
and, finally, the energy density,
\begin{equation}\label{3}
\epsilon =
 \sum_i g_i\int \frac{d^3p}{(2\pi)^3}E_i f_i,
\end{equation}
where,
\begin{equation}
f_i = \frac{1}{e^{(E_i-\mu_i)/T} \pm 1},
\end{equation}
with the upper (lower) signs being for Fermi-Dirac (Bose-Einstein)
statistics. The total density of baryons is calculated as in
(\ref{1}), but restricting the sum to  baryons $n_B$ and their
antiparticles $n_{\bar B}$. All of these results  correspond to the
grand-canonical formulation of the conservation laws. For central
heavy-ion collisions at energies below AGS, strangeness
conservation on average is not adequate and has to be implemented
exactly in terms of the canonical ensemble \cite{oeschler59}. A
more general result for  basic thermodynamic quantities in the
canonical formulation of strangeness conservation can be found in
Ref. \cite{review}.  The thermal particle phase-space of
hadronic  resonances  has also to be modified by including 
finite width \cite{review,suh}. This modification amounts to
convoluting the particle momentum distribution in Eqs.
(\ref{1}--\ref{3}) with the Breit-Wigner resonance form-factor.

The strange particle phase-space in Eqs.~(\ref{1}--\ref{3}) is
controlled by the $\gamma_s$ parameter \cite{r34}. The value of
$\gamma_s=1$ corresponds to a chemical equilibrium distribution,
whereas $\gamma_s\neq 1$ indicates a deviation from chemical
equilibrium. In addition,  Eqs. (\ref{1}--\ref{3}) are only
valid for point-like particles. Considering hadrons as extended
objects would require an implementation of interactions. This can
be done in the hard-core \cite{hardcore} and self-consistent
implementation \cite{gorenstein}. The first approach amounts to
the multiplication of all results in Eqs.~(\ref{1}),
(\ref{Entropy}) and (\ref{3}) by a factor,

\begin{equation}\label{4}
F(T,\mu_B,V_0)= \frac{1}{1+V_0n(T,\mu_B)},
\end{equation}
where the repulsive strength is determined  by the  volume parameter
$V_0$.

In the self-consistent implementation of the repulsive interaction
there is, in addition, a shift in chemical potential that
guarantees that all observables Eqs.~(\ref{1})-(\ref{3}) are obtained
as corresponding derivatives  from the partition function
\cite{gorenstein}. The explicit results and a detailed description
can be found in Ref. \cite{gorenstein}.

Before  applying the above results to  quantify    different
freeze-out criteria and comparing  their predications  with the
thermal model analysis of heavy-ion data, we first briefly
summarize basic concepts of these  criteria.

\subsection{Percolation Model}
Percolation theory was  successfully applied to the description of
critical properties of QCD matter \cite{hsp1}. In particular,
color deconfinement  in pure gauge theory  can be treated as a
percolation phenomenon \cite{hsp2}. The models based on percolation
theory also  provide a qualitative description of charmonium
production and suppression in heavy-ion collisions \cite{hsp3}.

In Ref.~\cite{satz}, percolation theory  was used to formulate and
quantify chemical freeze-out conditions in heavy-ion collisions.
It is assumed that, in the sector of vanishing baryon density,
hadronic matter freezes out according to the resonance gas
approximation and by vacuum percolation, while the region of
finite baryon density freezes out according to baryon percolation.
The condition that describes the freeze-out line in heavy-ion
collisions has been formulated  as \cite{satz}:
\begin{equation}\label{satz}
n(T,\mu) = \frac{1.24}{V_h}
\left[1-\frac{n_B(T,\mu)}{n(T,\mu)}\right]+\frac{0.34}{V_h}
\left[\frac{n_B(T,\mu)}{n(T,\mu)}\right].
\end{equation}
%which is what we have used in this analysis to test the dependence of this model on various assumptions
%and parameters.
The volume $V_h$ is  the hadronic size, which corresponds to a
radius of approximately $r_h\approx 0.8$ fm. The numbers 1.24 and
0.34 appearing in Eq.~(\ref{satz}) are calculated within
percolation theory~\cite{isichenko}. The smallest one is
determined   when the size of the largest cluster falls below the
size of the overall spatial volume. On the other hand, the number
1.24 is determined by the disappearance of any large-scale vacuum,
and  only the strongly interacting medium spans the entire space.

Eq.~(\ref{satz}), together with the results of the statistical-thermal 
model of the hadron resonance gas  on the  total density of
all hadrons $n(T,\mu_B)$ and baryon number density $n_B(T,\mu_B)$, 
provide a unique solution $T=T(\mu_B)$ that characterizes the
chemical freeze-out line in the $(T,\mu_B)$~--~plane within the percolation
model.
\subsection{Fixed Baryon+Anti-Baryon Density}
The contributions of baryons to thermodynamics is usually
characterized by  the net baryon density, i.e. the
difference between the density of baryons $n_B$ and anti-baryons
$n_{\bar B}$. In heavy-ion collisions, this  varies strongly with
collision energy. At SIS it is $\sim 1/3$ of nuclear matter
density $(\rho_0=0.17/\mathrm{fm}^3)$ and it
 strongly increases to reach a maximum at top AGS energies. It then
gently decreases up to RHIC, where it is only 
$\sim 1/10$ of $\rho_0$. Due to baryon number
conservation ($B=A_{part}$), the  variation of 
$n_B-n_{\bar B}=B/V$
 reflects the
corresponding change in the volume of the fireball. Indeed, a
comparison of the total yields of different particle species in
heavy-ion collisions with calculations in the statistical model,
or, alternatively, using  two pion correlations,  has shown that
the  volume has a minimum between AGS and the  lowest SPS energy
\cite{pbmv} and exhibits a dependence on $\sqrt s$ opposite to
that of $n_B$.

Despite  this large variation, it was first noticed in
Ref.~\cite{pbm} that the sum of baryon and anti-baryon densities, 
$n_B+n_{\bar B}$, remains remarkably constant.
 Thus, it was proposed  \cite{pbm} that in heavy-ion
collisions the chemical freeze-out curve can be obtained from the
condition,
%At low energies the net baryon density, i.e. baryons minus
%anti-baryons, is very high, while at high energies it is small due
%to the creation of large numbers of baryon--anti-baryon pairs.
%However, as first noticed in ref.~\cite{pbm}, the sum of baryon
%and anti-baryon densities remains remarkably constant over the
%whole energy range. The defining equation for the freeze-out curve
%is given by~\cite{pbm},
\begin{equation}\label{pbmv}
n_B+n_{\bar{B}} \simeq 0.12\,{\textrm{fm}}^{-3},
\end{equation}
of fixed density of total number of baryons and anti-baryons.
%  and gives an
%excellent description of the AGS, SPS and RHIC data but clearly
%overestimates the SIS energy region. Inclusion of nuclear
%fragments like deuterium in the hadronic gas improves the
%description at lower energies.
%

The observation \cite{pbmv}  that Eq.~(\ref{pbmv}) is 
satisfied in high energy heavy-ion collisions was important
phenomenologically to explain the equivalence at 160 and 40 AGeV
of the enhancement of the low mass dilepton yield observed by the
CERES Collaboration in Au-Pb collisions \cite{ceres} and recently
by NA60 in In--In reactions \cite{na60}. Theoretically, the
enhancement was argued \cite{rapp} to be due to rho meson
broadening through its rescattering with surrounding baryons.
Thus, the   broadening is determined by the total density of
baryons. The fact that in Au-Pb collisions at 160 and 40 AGeV the
total baryon  density is comparable, gave an argument for   the
observed similar structure of the low mass dilepton yield at these
energies.

Clearly, in low energy heavy-ion collisions between SIS and AGS, 
the production of anti-baryons is  suppressed. Thus, in this
energy range the net baryon number density, to a good
approximation, is equivalent to the  total density of baryons,
that is $n_B+n_{\bar B}$. Consequently, since the volume
at freeze-out  from SIS to AGS extracted from HBT correlations
\cite{pbmv} drops by a factor $1/3$, there should be a
corresponding increase in $ n_B+n_{\bar B}$. Because of the above,
the freeze-out condition (\ref{pbmv}) is not expected to
work in low energy heavy-ion collisions. The question remains as to at
what energy the violation of (\ref{pbmv})  as the condition of
particle freeze-out sets in. The answer to this  question requires
a quantitative analysis, which is given in the next section.

\subsection{Fixed Entropy Density over $T^3$}
The chemical freeze-out condition of fixed entropy density over
temperature to the third power,
\begin{equation}\label{entro}
\frac{s}{T^3} \simeq 7,
\end{equation}
was proposed independently from two very different considerations.
In Ref.~\cite{tawfik1,tawfik2}
it was used to extrapolate lattice gauge results from
 $\mu_B=0$  to finite values of $\mu_B$
by keeping $s/T^3$ fixed.
%This led to a very good description of the
%chemical freeze-out points.
 In Ref.~\cite{stankiewicz,horn} the entropy density was used to
separate between a baryon-dominant region and a meson-dominant one, 
in order to understand the rapid change in certain particle ratios
observed at lower SPS energies by the NA49
collaboration~\cite{NA49}. In the course of this analysis, it was
noted that, over a broad energy range,
%
% the whole freeze-out region
%(except in the SIS energy domain)
the total entropy density divided by $T^3$ remained constant, 
despite very large changes in the baryon chemical potential and
the temperature.

The freeze-out condition (\ref{entro}) is rather surprising. In the
ideal gas of  massless constituents  and for a net baryon free
system, the $s/T^3$ ratio describes the number of degrees of
freedom. In the case of massive particles  and at finite chemical
potential, this ratio  can be parameterized as $s/T^3\simeq
a(T,\mu_B)$. Thus,  for fixed $T$ and $\mu_B$, the  parameter  $
a(T,\mu_B)$ describes the effective number of degrees of freedom
in a system. The condition (\ref{entro}) tells us that this
effective number of degrees of freedom $a(T,\mu_B)$ is common for
all collision energies. In the thermodynamical context, this also
means that $a(T,\mu_B)$ is independent of thermal parameters at
chemical freeze-out. This is quite unexpected, as changing
the collision  energy implies not only changing the thermal
parameters but also the hadronic composition of the fireball.
Naively one would expect that in such a case the number of degrees
of freedom $a(T,\mu_B)$ should be a strongly varying function of
collision energy.
\subsection{Fixed $\eovern \simeq 1~\mathrm{GeV}$
}
This criterion was first proposed in Ref.~\cite{redlich1,redlich2}
as a link unifying results obtained at SIS with top  AGS and  SPS.
Since then, the description has proven its validity also at
RHIC~\cite{kaneta2,magestro} and at the lower SPS
energies~\cite{review,manninen,tounsi}. This  condition was
successfully used in the literature to make predictions
\cite{tounsi}  of freeze-out parameters at SPS energies of 40 and
80 AGeV  for Pb--Pb collisions long before the data were taken.
These predictions turned out to be in  agreement with the
statistical model analysis of recent experimental data obtained at
these collision energies \cite{manninen}.

The reason why  $\eovern$ is constant in heavy-ion collisions from
SIS up  to RHIC can be intuitively explained  as follows:  In a
non--relativistic system  $
 {{\langle E\rangle}/ {\langle N\rangle}}\simeq \langle m\rangle
+\frac{3}{2}T $, with $ \langle m\rangle$ being the average
mass in the thermal fireball \cite{hagedorn}. At SIS, the thermal mass
is of the order of the nucleon mass, whereas the freeze-out
temperature $T_f\simeq 50$ MeV. Thus, at SIS the  $\eovern$  ratio
is indeed comparable with 1 GeV. At  SPS and RHIC, the leading
particles in the final state  are pions. However, at chemical
freeze--out most of the pions are still hidden in the mesonic and
baryonic resonances. Thus, here the average thermal mass
corresponds approximately  to
%$\langle m
%\rangle\sim m_\rho$
the $\rho$--meson  mass. Consequently, with $T_f\sim$ 160--170 MeV
a comparable  value of $\langle E\rangle/\langle N\rangle$ is
obtained at SPS and RHIC as in a much lower energy at SIS.

We have already indicated that the chemical freeze-out line
dynamically coincides with the inelasticity line in heavy-ion
collisions. Thus, also  the $\langle E\rangle/\langle
N\rangle\simeq 1$ GeV condition should have the above  dynamical
interpretation. Recently, the correlation between the $\langle
E\rangle/\langle N\rangle\simeq 1$ GeV condition and inelasticity
has been investigated in central Pb--Pb collisions at the SPS  in
terms of the Ultra--relativistic Quantum Molecular Dynamics model
(UrQMD) \cite{urqmd}.
 A detailed
study has shown that there is a clear correlation between the
chemical break-up in terms of inelastic scattering rates and the
rapid decrease in energy per particle. If $\langle
E\rangle/\langle N\rangle$ approaches the value of  1 GeV the
inelastic scattering rates drop substantially and  further
evolution is due to elastic  collisions that preserve the chemical
composition of the collision fireball. Following  the above UrQMD
results, it is conceivable that  $\langle E\rangle/\langle
N\rangle \simeq 1$ GeV is related with inelasticity in heavy-ion
collisions. This could dynamically justify the applicability of
this freeze-out criterion.

\begin{figure*}[h]
\begin{center}
\includegraphics[width=12cm]{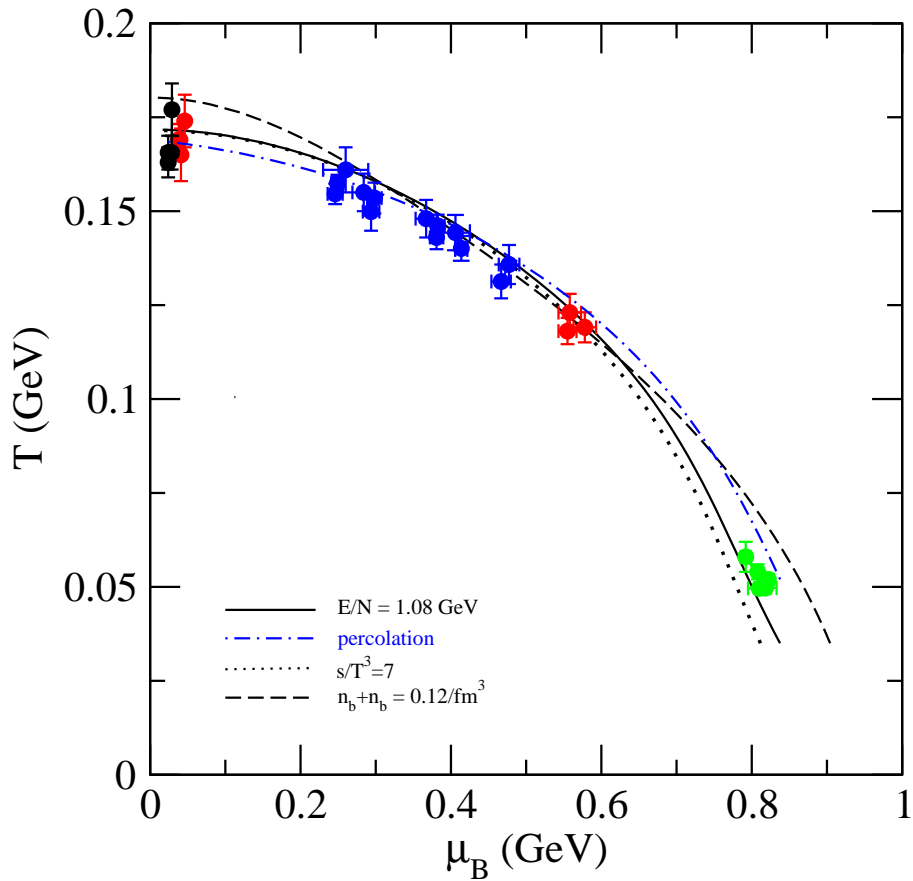}
\end{center}
\caption{
Description of chemical freeze-out by four different criteria discussed in the text.
}
\label{fo}
\end{figure*}

 It was noted recently in
Ref.~\cite{larissa} that the results from RHIC favor a higher
value for $\eovern$. This was interpreted as a change in the
baryonic composition of a two-component thermal source. We find that
a consistent description using $\eovern = 1.08$ GeV is quite
adequate and there is no need for different values at different energies.
The comparison of each of these freeze-out criteria with all known freeze-out parameters 
is shown in Figure~\ref{fo}.
%
%%%%%%%%%%%%%%%%%%%%%%%%%%%%%%%%%%%%%%%%%%%%%%
%%%%%%%%%%%%%%%%%%%%%%%%%%%%%%%%%%%%%%%%%%%%%%
\section{Sensitivity and robustness of freeze-out criteria}
The chemical freeze-out criteria discussed  in the last section are
not quite model independent. Their  predictions are sensitive to
various  assumptions and to the choice of  parameters. In this
section, we present a detailed discussion of the robustness of these
criteria. In this context, we consider the influence of the
repulsive hadronic interactions, strangeness suppression and 
hadron mass spectrum cut-off on the predictions of the different
freeze-out criteria.

\subsection{Sensitivity on excluded volume corrections}

The $\left<E\right>/\left<N\right>$ criterion involves the ratio
of two extensive quantities and, hence, is nearly independent of
excluded volume corrections. In the Boltzmann approximation, these corrections 
cancel exactly, provided that the same proper volume parameter is
applied. Otherwise, for example, pions and protons pick up different
correction factors through the shift in chemical potentials. In
quantum statistics, the shift in chemical potentials always leads
to a (small) change. The
energy per particle is also not strongly influenced if repulsive
interactions are implemented through a mean-field potential.
%%%%%%%%%%%%%%%%%%%%%%%%%%%%%%%%%%%%%%%%%%%%%%%%%%%%%%%%%
\begin{figure*}[t]
\begin{center}
\includegraphics[width=12cm]{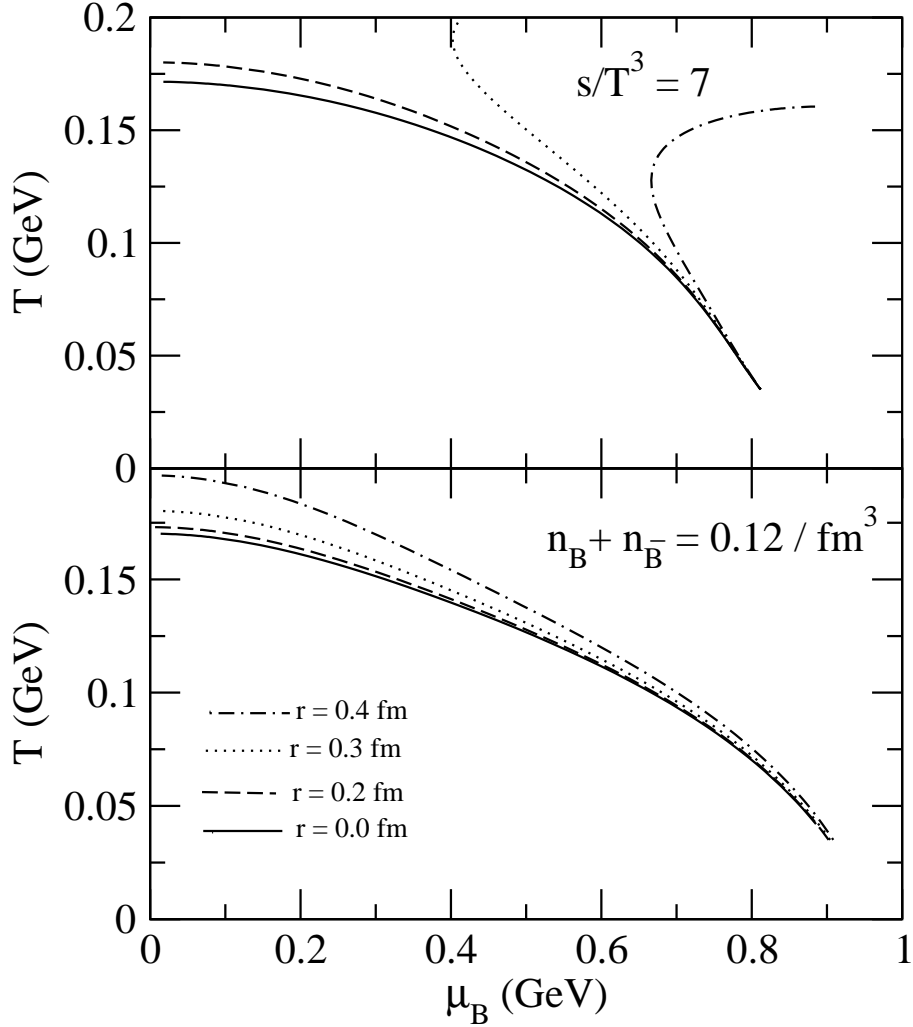}
\end{center}
\caption{
The effect of excluded volume corrections on the constant $n_B+n_{\bar{B}}$ (bottom)
and constant $s/T^3$ (top) freeze-out criterion.
}
\label{excluded}
\end{figure*}
The freeze-out  condition of fixed entropy density
divided by $T^3$, as
well as that of fixed total density of baryons plus anti-baryons, 
use extensive thermodynamic variables. Hence, 
the dependence on the excluded volume corrections is, in  general, 
not negligible. However, as it turns out, only a
 moderate dependence is found  for the
constant $n_B+n_{\bar{B}}$ criterion, which is shown in
Figure~\ref{excluded}. The repulsive strength of hadrons is
parameterized there by the radius parameter. The freeze-out curve
moves to higher values of $T$ as the radius  increases. A change
in radius of 0.1 fm easily leads to a change in temperature which
can be larger  than 10 MeV. The repulsive interactions 
suppress thermal particle phase-space. Thus, to reproduce the
same value of total density of baryons, one needs to increase
temperature at fixed $\mu_B$.

The percolation model criterion is  determined by the
total particle and the net baryon number density. It thus  shows a
similar dependence on excluded volume radius as discussed above.
This model is based on geometric considerations  which are
affected by volume parameters.  The changes are almost identical
to the previous case and are not shown explicitly.

Contrary to the percolation and fixed baryon density conditions, the
constant $s/T^3$ criterion shows a wild dependence on the  excluded
volume radius (as seen in Figure~\ref{excluded}). This  casts
considerable doubt on the value of this criterion for model
building.  It is clear that only the complete neglect of excluded
volume corrections leads to a realistic freeze-out curve. It is
well known, however, that such corrections cannot be ignored,
especially at low temperatures, as this leads e.g. to  a totally
unrealistic picture  for the phase transition \cite{esko} and is
also inconsistent with the properties of the nuclear potential.
\subsection{Sensitivity on resonance spectrum mass cut-off}
\begin{figure*}[t]
\begin{center}
\includegraphics[width=12cm]{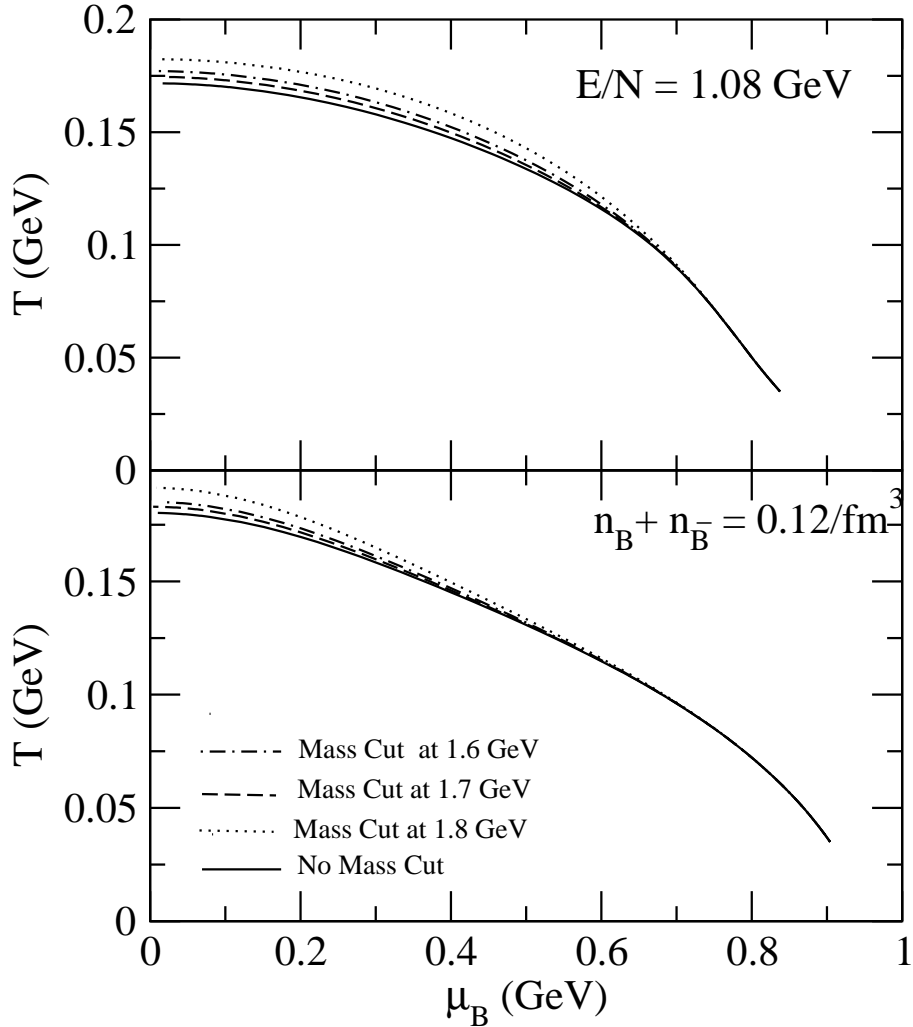}
\end{center}
\caption{
The effect of the resonance spectrum mass cut-off on the
chemical freeze-out curve: top constant $\left<E\right>/\left<N\right>$, bottom
 $n_B+n_{\bar{B}}$ freeze-out criterion.
}
\label{resonance}
\end{figure*}

The thermal-statistical model relies very strongly on input from
the particle data table \cite{PDG}. Heavy resonances are, in general, 
suppressed by the Boltzmann factor $e^{-E/T}$. However, when the
temperature or baryon chemical potential increase, the
contribution of heavy resonances also increases and this becomes
especially noteworthy in the small $\mu_B$ -- high $T$ region. As
the information about heavy resonances is fragmentary, several
analyses have introduced a cut-off in masses, such that resonances
heavier than this cut-off are not taken into account in the
thermal-statistical model. The effects of this mass cut-off are
similar for all criteria considered. As an example, we show the
influence on the $\left<E\right>/\left<N\right>$ = 1 GeV criterion
 and on the finite density
$n_B+n_{\bar{B}}$ = 0.12/fm$^3$ criterion in
Figure~\ref{resonance}. It is seen in Figure~\ref{resonance}
that the largest influence of the mass cut in the hadron spectrum
appears at high temperature and low $\mu_B$. At fixed 
$\mu_B$,  the changes in the freeze-out  temperature never
exceed 10 MeV.
%%%%%%%%%%%%%%%%%%%%%%%%%%%%%%%%%%%%%%%%%%%%%%%%%%%%%%%%%%%%%%%%
%%%%%%%%%%%%%%%%%%%%%%%%%%%%%%%%%%%%%%%%%%%%%%

The cut-off on the hadronic mass spectrum
in the thermal-statistical model is particularly relevant when calculating
particle yields. This is because the decay properties of heavier
resonances into lighter hadrons are needed and are  not always well
established. However, for the description of  basic
thermodynamic observables  characterizing the properties of the
hadronic  medium, like the energy, entropy, particle and baryon number
density, the only resonance characteristics needed are the masses,
degeneracy factors and quantum numbers. In most cases, 
this information is known even for heavy mesonic  and baryonic
states.

\subsection{Sensitivity on strangeness suppression}

Strange particle yields, their ratios and momentum distributions
are  relevant observables that  characterize the origin,
composition and production dynamics in heavy-ion
collisions \cite{review,r34}. However, their contribution to
global thermodynamic observables, summarized in Eqs.
(\ref{1}-\ref{3}), is not essential. In low energy heavy-ion
collisions at SIS up to AGS, strange particles are only rarely produced
and can be almost completely neglected when considering   global
thermodynamic characteristics of the collision fireball. At  SPS
and RHIC, due to the much higher temperature, the strangeness degrees
of freedom contribute about  $25\%$
 to the total particle density. Due to this, 
one should not  expect very large changes in the predictions of
different freeze-out criteria when modifying the strange particle
content.

In the statistical model under consideration, a modification of
the strange particles from chemical equilibrium is described by
the  $\gamma_s$ factor  or by the canonical
constraints imposed by exact strangeness conservation.

\begin{figure*}[h]
\begin{center}
\includegraphics[width=12cm]{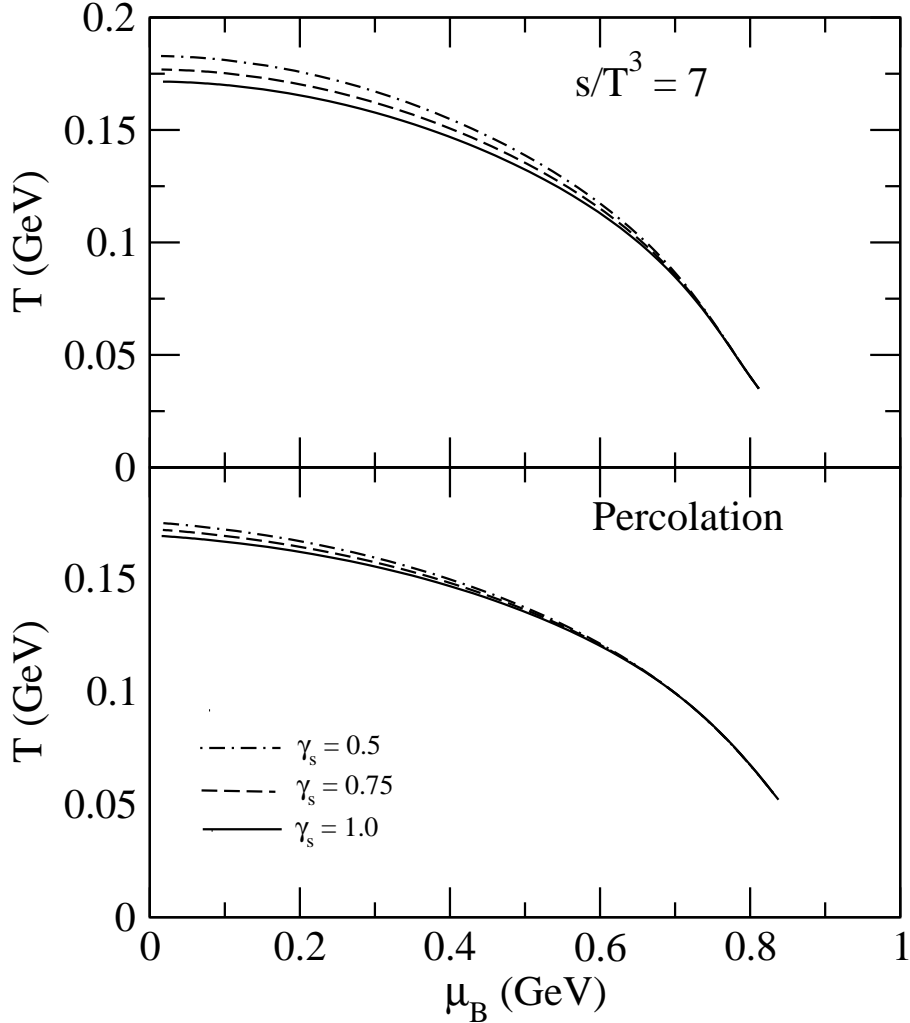}
\end{center}
\caption{
The effect of varying $\gamma_s$ on the constant $s/T^3$ (top)
and the percolation criteria (bottom).
}
\label{gammas}
\end{figure*}

The strangeness-suppression (or enhancement) factor $\gamma_s$
modifies the multiplicities of strange and anti-strange  hadrons,
parameterizing a deviation from chemical equilibrium, in the 
Boltzmann approximation, in the manner
given by the equation below,
\begin{equation}
\left<n_i\right> = \gamma_s^{n_s}\left<n_i\right>_{\mathrm{equilibrium}}.
%\left<n_i\right> =g_i\int\frac{d^3p}{(2\pi)^3}
%\frac{1}{\gamma_s^{-n_s}e^{(E-\mu)/T}\pm 1} .
\end{equation}
It  does not affect the non-strange 
hadrons except for those contributions coming  from decays of 
strange resonances. 
Its
influence on the chemical freeze-out curve is therefore small and,
at low energies, negligible. This is shown for the $s/T^3$
criterion and for the percolation model in
Figure~\ref{gammas}. The effect on the other freeze-out criteria
is similar and not shown explicitly.
%%%%%%%%%%%%%%%%%%%%%%%%%%%%%%%%%%%%%%%%%%%%%%
%%%%%%%%%%%%%%%%%%%%%%%%%%%%%%%%%%%%%%%%%%%%%%

\begin{figure*}
\begin{center}
\includegraphics[width=12cm]{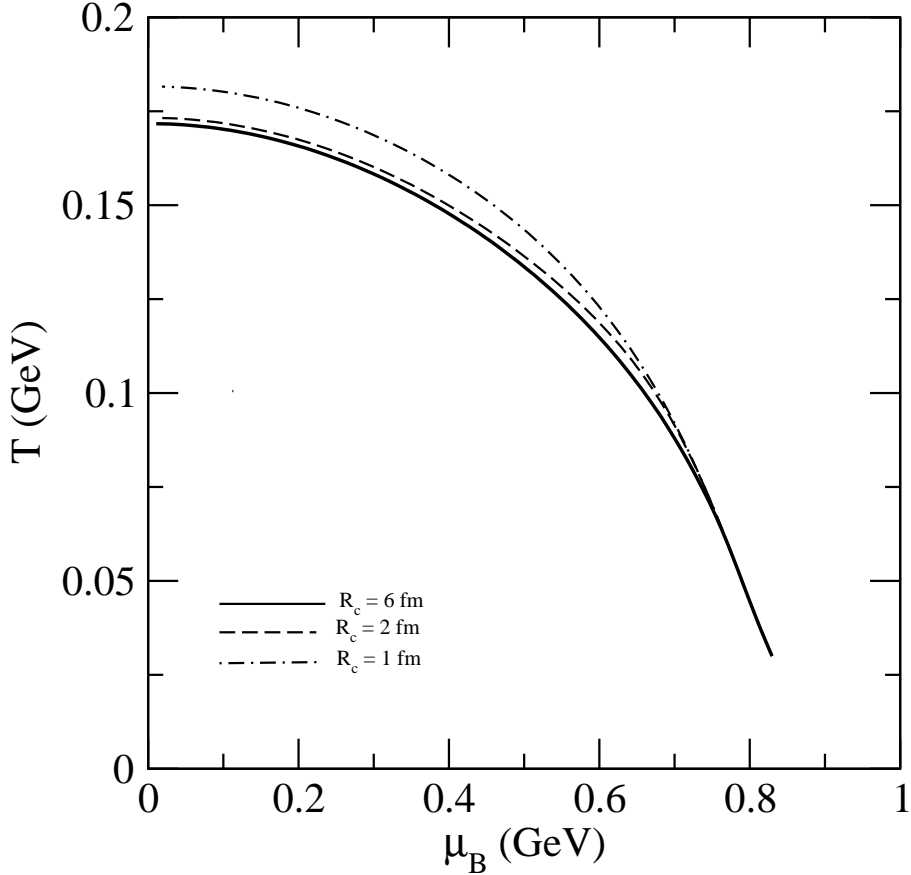}
\end{center}
\caption{
The effect of the canonical radius on the constant $s/T^3$ freeze-out criterion.
}
\label{canonic}
\end{figure*}

For small systems, various
corrections become important. In our
considerations, these affect mainly
the production  of
strange particles
in heavy-ion collisions at low energies due to
the very small number of produced  particles. In this case, strangeness
conservation should be treated exactly, which results in  so-called
canonical corrections. These depend on the volume over which
strangeness is balanced. This  correlation volume does
not need to be as large as the volume of the overall system. Several
calculations have been made where strangeness conservation is
imposed over volumes which are smaller than the overall volume.
These are parameterized by a canonical radius. In
Figure~\ref{canonic}, one typical example shows  the
influence of this canonical radius on the freeze-out criteria.
%%%%%%%%%%%%%%%%%%%%%%%%%%%%%%%%%%%%%%%%%%%%%%%%%%
%%%%%%%%%%%%%%%%%%%%%%%%%%%%%%%%%%%%%%%%%%%%%%
Similarly as for the strangeness suppression factor,
 since  the relative number of strange particles is small, the
effect of the canonical radius on the chemical freeze-out curve is
small. All criteria are only weakly dependent on variations
in the canonical parameter. Small differences  are only seen above
AGS energies, where the contribution of the strangeness sector is not
negligible. At low energy, this effect is entirely irrelevant.
However, for the discussion of strange particles or
their yields, the influence of canonical effects is essential
\cite{oeschler59,tounsi,tounsi1,ingrid}. This is particularly the
case for multistrange baryons in low energy central collisions, as well as in
peripheral A--A collisions at higher energies
\cite{tounsi,tounsi1}. The $\gamma_s$ factor has a similar
influence on strange and multistrange particle yields
\cite{re2,manninen}.

Comparing different freeze-out criteria, we have been bound by the
knowledge of the lowest chemical freeze-out parameters  at SIS
energy. Clearly, the thermal-statistical model considered here is limited
and cannot be used if the assumption of energy conservation
on the average is not valid. This indeed is the case in low energy
heavy-ion collisions, where the  full micro-canonical formulation
\cite{becm} of the partition function is required. This also means
that the extrapolation of different freeze-out criteria to the $T\simeq
0$ limit and within the grand-canonical formulation is
questionable. Until now the model has shown its validity in the
description of heavy-ion data up to
beam energies of about 800 MeV.

There is  a  class of further criteria (e.g. the constant
energy or total particle number density) that are consistent with
the freeze-out parameters in the energy range above the lowest SPS energies.
However, since these conditions are based on intensive observables, 
they suffer from variations with the parameterization of
repulsive interactions. In addition, both of these criteria are
inconsistent with the SIS freeze-out parameters. 
\section{Summary and Conclusions}

From the results presented above one can conclude that, on a
phenomenological level, the various freeze-out criteria give quite
satisfactory  descriptions of the particle multiplicities measured
in heavy-ion collisions. Deviations can be reduced by suitably
modifying input parameters in the different models. However, it is
clear that some of the criteria are not very stable or robust when
one applies small changes to the input considerations and are, therefore, 
more model dependent.

A further
method to  differentiate freeze-out conditions
would be to calculate particle excitation functions. The
differences between such criteria would be even more transparent, 
as some particle excitation functions are very sensitive to the
value of the thermal parameters. A good example is the $K/\pi$ ratio
which at SIS energy is a very sensitive probe of temperature and
canonical suppression effects. The energy per particle criterion
has been  shown to provide a consistent  description of
the  particle excitation functions over the whole energy range.
Some of the results of this criterion, like the energy dependence
of the $\Lambda /\pi$ and $K^+/K^-$ ratios for different
centrality and
collision energies, have been confirmed  by recent heavy-ion
data.

%A summary of the sensitivity
%on model assumptions and parameters  is presented  in Table~\ref{tab:comparison}.
%Also shown in this Table is the   estimate of the quality of the
%criteria by calculating the value of the $\chi^2$ for each case.
The main differences between the  chemical
freeze-out criteria analyzed here appear in  the region of low temperature and
large  baryon chemical potential, that is for SIS up to AGS
energy. This region obviously favors  the $\eovern$ criterion as
the most appropriate one to describe chemical freeze-out over all
energies. This is supported quantitatively by the fact that this
freeze-out condition  has the lowest $\chi^2$ per degree of
freedom of all the criteria considered in this analysis and was
shown to provide a good description of  different  particle
excitation functions in heavy-ion collisions. The energy per
particle is also consistent through transport model calculations
with the dynamical interpretation of chemical freeze-out as the
inelasticity condition  in heavy-ion collisions.

\section*{Acknowledgments}
Two of us (J.C. and S.W.) would like to thank the theory division
of the GSI for their hospitality.  The  partial  support of the
Polish Committee for Scientific Research under contract 2P03
(06925) and the Polish--South--African Research Project is
acknowledged. J.C. acknowledges the financial support of
the Alexander von Humboldt Foundation and the Italy-South Africa
co-operation programme and discussions with F. Becattini. 
K.R also  acknowledges
stimulating discussions with P. Braun-Munzinger, H. Satz  and J.
Stachel.

%\begin{thebibliography}{50}

\end{document}